\documentstyle[preprint,aps]{revtex}
\tighten
\preprint{\vbox{Submitted to Physical Review {\bf C}
                \hfill IU/NTC 96-02}}
\begin{document}

\title{Detection of Atmospheric Neutrinos and \\ Relativistic  
Nuclear Structure Effects}

\author{Hungchong Kim~\footnote{E-mail:~hung@phya.yonsei.ac.kr}
$^{1,2}$, S. Schramm~\footnote{E-mail:~schramm@tpri6e.gsi.de}
 $^3$, C. J. Horowitz
~\footnote{E-mail:~charlie@proteus.iucf.indiana.edu} $^1$}
\address{$^1$ Nuclear Theory Center, Indiana University,
Bloomington, Indiana 47408, USA \\
$^2$ Department of Physics, 
Yonsei University, Seoul, 120-749, Korea \\
$^3$ GSI, D-64220 Darmstadt, Germany} 
\date{\today}
\maketitle
\begin{abstract}
Neutrino-nucleus cross sections for the detection of atmospheric
neutrinos are calculated in relativistic impulse and random phase
approximations.  Pion production is estimated via inclusive $\Delta$-
hole nuclear excitations.  The number of pion events can confuse the
identification of the neutrino flavor.  
As a further source of pions we
calculate coherent pion production (where the nucleus remains in the
ground state).  We examine how these nuclear structure effects
influence atmospheric  neutrino experiments and study possible 
improvements in the present detector simulations.

\end{abstract}
\eject
\narrowtext
\section{Introduction}  
\label{sec:cint}                                                
                                           
Aside from the puzzling solar neutrino data there exist extensive
measurements of atmospheric neutrinos, which
are produced in cosmic ray interactions with the earth's atmosphere.
Cosmic rays, mostly protons or $\alpha$ particles, hitting
the atmosphere produce pions, which undergo the following
decays:
\begin{eqnarray}
\pi^{\pm}&\rightarrow &\mu^{\pm}+\nu_{\mu}\ (\bar{\nu}_{\mu})
\nonumber\\
& &\mu^{\pm}\rightarrow e^{\pm}+\nu_{e}\ (\bar{\nu}_{e})+
\bar{\nu}_{\mu}\ (\nu_{\mu})\nonumber\ .
\end{eqnarray}
From this, one expects the neutrino flavor ratio,
\begin{eqnarray}
r= {\nu_\mu +{\bar \nu_\mu} \over \nu_e +{\bar \nu_e}}\ ,
\label{rat}
\end{eqnarray}
to be about 2.  However, two existing experiments,
IMB~\cite{imb} and Kamiokande~\cite{kamiokande}, 
consistently measured a 
ratio of about 1 by studying 
charged-current neutrino-nucleus reactions [($\nu_e$, $e$) and
($\nu_\mu$, $\mu$)] in large underground
water detectors.  This could mean either there is a depletion of 
muon-type neutrinos or an enhancement of electron-type neutrinos.
While the separate $\nu_\mu$ and
$\nu_e$ neutrino fluxes differ in various calculations, 
their ratio, however,
seems to be largely model-independent~\cite{eng}, and 
the theoretical results are in clear contradiction with data. 
This discrepancy is known as the ``atmospheric neutrino anomaly''.

One explanation of this result could be the existence
of neutrino oscillations of the muon neutrino into 
some other neutrino
species~\cite{solar} which would reduce the $\nu_\mu$ flux. 
However, before 
drawing this conclusion, one should also investigate more 
conventional sources
for uncertainties in the measured flux ratio, which originate from
nuclear physics and  detector specifics.
One crucial point is the experimental separation of   
electron and muon events.
Most  data  are measured using water detectors. In a charged-current  
weak interaction with an $^{16}$O nucleus, the incoming neutrino
generates an outgoing lepton which is detected through its
\v{C}erenkov radiation. Electrons and muons are distinguished by the
characteristics of their tracks. Electrons produce a showering
``fuzzy'' track, whereas muons are identified by their nonshowering
track.

Pions constitute a major background in this way of determining 
the neutrino flavors. In the experimental analysis, two- and
more ring events, which signal the production of several charged 
particles, are rejected. However, a charged pion
can be confused with a muon event if the lepton energy is below
\v{C}erenkov radiation threshold, and a neutral pion, with 
one decay photon missing,
can be counted as an electron event.  Therefore, 
a reasonable determination of the flavor ratio can only be 
achieved when the pion events are properly taken into account.

One major source of pions is the decay of a $\Delta$ produced 
in the neutrino scattering. The delta can 
subsequently decay into a pion and
a nucleon. An estimate of the number of pion events is only 
possible when the $\Delta$-h response is properly calculated.
Indeed, Monte Carlo simulation of the Kamiokande 
experiment~\cite{monte} estimated the number of pion events based 
on a rather complicated 
nonrelativistic model~\cite{fogli}.
However, atmospheric neutrinos have a broad energy spectrum 
ranging
from a few hundred MeV to several GeV.  
It is therefore crucial to have a relativistic formalism that allows 
the calculation of the cross section for arbitrary kinematics.

As another source of pions, we consider coherent pion 
production and its role in the detection of atmospheric neutrinos.
Coherent pions are produced in both charged- and neutral-current 
neutrino-nucleus interactions which leave the nucleus in its 
ground state.  
Traditionally, the coherent $\pi^0$'s are distinguished from the 
resonantly produced pions due to their strongly 
forward peaked angular 
distribution.  However, in atmospheric neutrino experiments, due 
to a lack of
directional information of the incoming neutrinos, it is not 
possible to
distinguish the coherent $\pi^0$ events from incoherent pions.  

In the case of coherent charged pions the forward angle dominance 
of the cross section leads to a small opening angle between two 
charged particles ($\mu$ or $e$ and charged pion) in the final state, 
and these two particles could be detected as one isolated 
electromagnetic shower.  In contrast to
incoherent charged pions whose \v{C}erenkov tracks are
usually identified as muon events, the special characteristics 
of coherent pions 
could lead to an increase of identified $\nu_e$ events
and thus to underestimating  the $\nu_\mu$ to $\nu_e$ ratio. 

This paper is organized as follows.  In Sec.~\ref{sec:incl} 
we present 
the formalism for neutrino-nucleus cross section including p-h and 
$\Delta$-h
nuclear excitations within the framework of
a relativistic mean-field model of the nucleus. 
Sec.~\ref{sec:coh_for} contains  
the formalism for coherent charged and neutral pion  
production. 
The resulting cross sections and production rates are shown in 
Sec.~\ref{sec:result}. We close with a discussion  and outlook in 
Sec.~\ref{sec:con}.
 

\section{ Inclusive Cross Section}
\label{sec:incl}

We outline the formalism for inclusive neutrino-nucleus scattering 
cross section
including p-h (nucleon particle-hole) and $\Delta$-h (delta-hole) 
nuclear excitations.  We consider a 
neutrino with four-momentum $k=(E_\nu, {\bf k})$ which scatters from 
a nucleus 
via $W^{\pm}$ boson exchange producing a charged lepton with 
four-momentum
$k'=(E_{\bf k'}, {\bf k'})$.  Using impulse approximation, 
the double differential scattering cross section from  a target 
nucleus with
mass number $A$ is 
given by (we assume a symmetric $N=Z$ nucleus):
\begin{eqnarray}
{d^2\sigma\over {d\Omega _{\bf k'} d E_{\bf k'}}}=- {A G_F^2\,
{\rm cos^2\theta _c}
\,|{\bf k}'|\over
{32 \pi^3 \rho E_\nu}}\,{{\rm Im}\,[L_{\mu\nu}\ \Pi_A^{\mu\nu}]}\ . 
\label{dcross}
\end{eqnarray}
Here $\rho=2k_F^3/3\pi^2$ is the baryon density with Fermi 
momentum $k_F$, 
$\theta_c$ the Cabbibo angle (${\rm cos}^2\theta_c=0.95$), $G_F$ 
is the Fermi 
constant.
The leptonic tensor $L_{\mu\nu}$ is defined as
\begin{equation}
L_{\mu\nu} = 8 \left( k_\mu k'_\nu + k_\nu k'_\mu -
k \cdot k' g_{\mu\nu} \mp i \epsilon_{\alpha\beta\mu\nu}
k^{\alpha} k'^{\beta} \right)\ ,
\label{lep}
\end{equation}
with the minus (plus) sign denoting neutrino (anti-neutrino) 
scattering.
$\Pi^{\mu\nu}_A$ is the polarization tensor of the target nucleus 
for the
charged weak current.  Here we consider p-h, $\Pi^{\mu\nu}_{ph}$, 
and 
$\Delta$-h, $\Pi^{\mu\nu}_{\Delta h}$, contributions to the 
polarization:
\begin{equation}
\Pi^{\mu\nu}_A=\Pi^{\mu\nu}_{ph}+\Pi^{\mu\nu}_{\Delta h}\ .
\end{equation}
In the impulse approximation, the p-h polarization takes a simple
form,
\begin{equation}
i\Pi^{\mu \nu}_{ph} = \int{d^4p\over (2\pi)^4}\, 
  {\rm Tr}[G(p+q)\,\Gamma^\mu \,G(p)\,\Gamma^\nu] \;, 
\label{tun} 
\end{equation}
with the weak-interaction vertex given in terms of single-nucleon
form factors parameterized from on-shell data,
\begin{eqnarray}
\Gamma^\mu(q^2)&=& F_1(q^2)\gamma^\mu +
              iF_2(q^2) \sigma^{\mu\nu} {q_\nu \over 2M} - 
               G_A(q^2)\gamma^\mu \gamma^5 + 
               F_p(q^2)q^\mu\gamma^5\nonumber\ ,\\ 
&&\quad (q^2 \equiv q_{0}^{2}-{\bf q}^{2})
\label{cur}
\end{eqnarray}
The form factors $F_1, F_2, G_A$ and $F_p$ are given in the 
appendix of 
Ref.~\cite{hung}.  The
pseudoscalar form factor $F_p$ is constructed from PCAC, and its 
contribution 
is suppressed by the small lepton mass.  

In an 
approximation where the nucleus is in a mean-field ground state of 
the Walecka model, 
the nucleon propagator $G(p)$
is given by~\cite{brian}
\begin{eqnarray}
 G^*(p)=(\not\!{p}^{*} + M^*) 
 \biggl[{1\over p^{*2}-M^{*2} +i\epsilon} + 
 {i\pi \over E^*_{\bf p}} \delta({p}_{0}^{*}-E^{*}_{\bf p})
 {\rm \theta}(k_F-|{\bf p}|)\biggr] \label{gmf} \;, 
\end{eqnarray}
where the effective mass  $M^*$ and energy $E_p^*$ are shifted from 
their
free-space value by the scalar ($S$) and timelike component 
($V$) of the mean fields,
\begin{equation}
 M^*=M+S \;; \quad E^{*}_{\bf p} =\sqrt{{\bf p^2}+M^{*2}} \;; \quad
 {p}^{*\mu}=(p^0-V,{\bf p}) \;.
\end{equation}
In the Fermi gas approximation of non-interacting particles, 
$V$ and $S$ 
are set to zero.
Analytic expressions for the imaginary part of 
the polarization [Eq.~(\ref{tun})] can be found in Ref.~\cite{hung}.   

The impulse approximation can be improved
by  including RPA effects.  An RPA
calculation uses the same expression for the cross section as in 
Eq.~(\ref{dcross}) with the replacement:
\begin{equation}
\Pi^{\mu\nu} \rightarrow \Pi^{\mu\nu}_{RPA}= 
\Pi ^{\mu\nu}+\Delta\Pi_{RPA}^{\mu\nu} \;.
\end{equation}
$\Delta\Pi_{RPA}^{\mu\nu}$ represents many-body correlations mediated 
by isovector particles, $\pi$ and $\rho$ incorporated with the 
Landau-Migdal parameter $g'$.
 
The cross section involving $\Delta$-h excitations is calculated 
similarly 
in Hartree approximation~\cite{delta} using the $\Delta$-h 
polarization 
$\Pi_{\Delta h}^{\mu\nu}$. 
The weak interaction contains vector current ($v$) and 
axial-vector current ($a$) contributions,
therefore, we split $\Pi_{\Delta h}^{\mu\nu}$ into
\begin{equation}                                
\Pi^{\mu\nu}_{\Delta h} = (\Pi_{\Delta h}^{vv})^{\mu\nu} 
+ (\Pi_{\Delta h}^{aa})^{\mu\nu}       
+ (\Pi_{\Delta h}^{va})^{\mu\nu}            
+ (\Pi_{\Delta h}^{av})^{\mu\nu}\ . 
\end{equation}
Here $(\Pi_{\Delta h}^{va})^{\mu\nu}$ and
$(\Pi_{\Delta h}^{av})^{\mu\nu}$
are interference terms of the vector and axial-vector
currents.  These polarization tensors are written in terms of the 
spin $3/2$
delta propagator and appropriate weak vertices as,
\begin{eqnarray}                                                        
(\Pi_{\Delta h }^{ij})_{\mu\nu} =&-& i \int{d^4p\over (2\pi)^4}\,
{\rm Tr}[\Gamma^i_{\beta\mu}(-q,-p)\ S^{\beta\alpha}(p)\                   
\Gamma^j_{\alpha\nu}(q,p)\  G(p-q)\ ]\nonumber \\                         
&+& (q_\mu \rightarrow -q_\mu)\ ~~~~~~~~~~(i,j) = (a,v)\ . 
\label{eq:pol}             
\end{eqnarray}                                                         
The trace is taken for the Dirac matrices as well as the isospin 
matrices.
For $S^{\mu\nu}(t)$ we take the Rarita-Schwinger
form of the free spin 3/2 propagator with  momentum
$t$ :
\begin{equation}
S^{\mu\nu}(t)=-{\not\!t + M_\Delta \over t^2 - M_\Delta^2+i\epsilon} 
\biggr
[ g^{\mu\nu}
-{1 \over 3} \gamma^\mu \gamma^\nu-{2 \over 3} 
{t^\mu t^\nu \over M_\Delta^2} +
{t^\mu \gamma^\nu - t^\nu \gamma^\mu \over 3 M_\Delta } \biggl ]\ .
\end{equation}
The vector part of the nucleon-delta vertex has been studied in the 
case of the
$\gamma N \Delta$ transition~\cite{wehr89},
\begin{eqnarray}
\Gamma^{V}_{\mu\nu} (q,p) &&= \sqrt{2} F^\Delta(Q^2) T^{\pm} \biggr
[ (-q_\mu \gamma_\nu+g_{\mu\nu}\not\!q) M_\Delta \gamma_5 +
(q_\mu p_\nu -q \cdot p g_{\mu\nu}) \gamma_5 \biggl]\nonumber \\
&&\equiv \sqrt{2} F^\Delta T^{\pm} V_{\mu\nu}\ 
\end{eqnarray}
with the isospin raising or lowering operator $T^{\pm}$ \cite{delta}. 
The vertex for the axial $N\Delta$ transition is given 
by~\cite{tjon,ann,towner}
\begin{eqnarray}
\Gamma^A_{\mu\nu}&=&-{r_{N \Delta} \over \sqrt{2} }
G_A T^{\pm}
(g_{\mu\nu}-{\gamma_\mu \gamma_\nu \over 4})\ 
\end{eqnarray}
with $r_{N\Delta} = f_{\pi N\Delta}/f_{\pi NN} \sim 2$.
The form factors $F^\Delta$ and $G_A$ are defined in the appendix of 
Ref.~\cite{delta}.

In mean-field approximation, the propagation of a $\Delta$ is modified
by background scalar and vector fields similarly as the nucleon 
propagator.  
This can be understood from a chiral soliton model where the delta
is a rotational excitation of the nucleon. That is, in the medium, 
the delta is expected to be influenced by the same magnitude of the 
vector and 
scalar potentials as the nucleon's because $\sigma$ and $\omega$ are 
isoscalar. 
The calculation of the cross section 
proceeds in the same way as in the case of the free $\Delta$.    

To include the delta decay width in our calculation, we average the 
nuclear
response over the delta mass with a Breit-Wigner 
distribution~\cite{wehr89}.  
The averaged cross section follows as
\begin{eqnarray}
<{d^2\sigma\over {d\Omega _{k'} d E_{\bf k'}}}> &\sim&
\int d\mu^2 {d^2\sigma\over {d\Omega _{k'} d E_{\bf k'}}} (\mu)
f(\mu^2) ~/~ \int d\mu^2 f(\mu^2)~~~~, \\
f(\mu^2) &=&
{M_\Delta \Gamma_\Delta \over (M_\Delta^2-\mu^2)^2 + 
M_\Delta^2 \Gamma_\Delta^2 }\
\end{eqnarray}      
integrating from threshold to infinity.
However,
the decay width of a delta in the nuclear medium is not well 
determined 
both theoretically and experimentally.   The $\pi$N decay 
channel of the delta is partially blocked by Pauli blocking but 
there is
an additional spreading width.  Thus, determining the in-medium 
decay width
is a nontrivial problem.  
Here, we assume the in-medium width
to be  the same as the free width $\Gamma_{\Delta}=115$ MeV. 
More theoretical and experimental work has to be done regarding this
question, however.

\section{ Coherent Pion Production}
\label{sec:coh_for}

Coherent pions are produced as decay products of virtual p-h or 
$\Delta$-h excitations of a finite nucleus.  More specifically,
the momentum transfer from the incoming neutrino virtually excites 
the nucleus through p-h or $\Delta$-h and the nucleus decays back to
its ground state by emitting a pion. Here we derive the formalism for
the $\Delta$-h excitation  as it dominates 
the p-h excitations of the nucleus. The corresponding scattering 
diagram is
shown in Fig.~\ref{cohfig}.    
Note that the coherent pion is on-shell even though the momentum 
transfer in the initial scattering is space-like.   
The missing momentum is provided
by the recoil of the nucleus.

In a local density approximation, the finite-nucleus polarization 
$\Pi_{\rm FN}$$(q,q';\omega)$ with incoming and outgoing momenta
$q$ and $q'$, respectively, is approximated in terms of the nuclear 
matter 
polarization $\Pi_{\rm NM}$$(q,q';\omega)$, 
\begin{equation}
\Pi^{\mu}_{\rm FN}(q,q';\omega) = 
{\rm V}\ {\rm F}(|{\bf q} -{\bf q'}|)\ 
\Pi^{\mu}_{\rm NM}(q,q';\omega)~~ ,
\end{equation}
where ${\rm F}(|{\bf q} -{\bf q'}|)$ is the elastic 
form factor of the target nucleus. We obtain 
${\rm F}(|{\bf q} -{\bf q'}|)$
by Fourier transforming the nuclear ground state density 
obtained from a self-consistent relativistic mean
field calculation for finite nuclei~\cite{horo81}.
The volume for a target nucleus with mass number A is given by 
$${\rm V}={3 \pi^2 {\rm A} \over 2 k_F^3}\ .$$ 

As shown in Fig.~\ref{cohfig}, the incoming neutrino interacts 
weakly
with the nucleus exciting an intermediate $\Delta$-h state.
The vertex of the weak interaction contains  
vector and axial-vector parts while 
the $\pi N\Delta$ vertex, on the other hand, 
involves only the 
axial-vector part contracted with a pion four-momentum $q'$.  
Since the coherent-pion production cross section is dominated by 
pion momenta ${\bf q'}$  
parallel to the momentum transfer ${\bf q}$, 
there is negligible contribution from the part of the polarization 
that mixes the axial-vector and vector current.
We ignore this small contribution throughout this paper.
Consequently, the cross section for coherent pion production is 
identical for neutrinos and antineutrinos.

The double differential cross section for coherent pion production is 
\begin{eqnarray}
{d^2\sigma\over d E_\pi d\Omega_{\bf k'}}&=&
{|{\bf k'}| |{\bf q'}|\ G^2_F\ {\rm V}^2 \over
256 \pi^4 E_\nu}
\int^1_{-1} d {\rm cos} \theta_{q'}\ {\rm F}^2({\rm cos} \theta_{q'})
\biggr[ L_{00} |\Pi_{\rm NM}^{0}|^2\nonumber \\
&& + 2 L_{01}\ {\rm Re}[\Pi_{\rm NM}^{0} (\Pi^{1}_{\rm NM})^*]
+L_{11}|\Pi_{\rm NM}^{1}|^2 \biggl]\label{double}\ ,
\end{eqnarray}
where $\theta_{q'}$ is the angle between ${\bf q}$ and ${\bf q'}$.
Here we observed that the integral in Eq.~(\ref{double})
is dominated by small angles, which implies that
contributions from transverse
components of the polarization, $\Pi_{\rm NM}^{2}$ and 
$\Pi_{\rm NM}^{3}$,
are suppressed. Energy conservation requires
the pion energy $E_\pi$ to be equal to the energy transfer to the
nucleus $q_0$.
The pion three-momentum $\bf{q'}$ is obtained from the mass-shell 
condition
\begin{equation}
|{\bf q'}|=\sqrt{q_0^2-m_\pi^2}\ .
\end{equation}
The antisymmetric part of the leptonic tensor $L_{\mu\nu}$ 
does not contribute to
Eq.~(\ref{double}), as it is contracted with the vector axial-vector
mixing term of the polarization.  The nuclear matter polarization, 
$\Pi_{\rm NM}^\mu$, can be
evaluated for coherent charged or neutral pion production by taking
appropriate vertices.  
The calculation of the real and imaginary parts of
the polarizations $\Pi_{\rm NM}^0$ and $\Pi_{\rm NM}^1$ is summarized 
in Appendix.

In a relativistic mean field approximation,
the properties of the nucleon and $\Delta$ are modified
by strong scalar and vector fields in the
nuclear medium. These effects can be incorporated by
modifying
the nucleon and delta propagators as we described in the 
inclusive calculation.
Since the axial-vector vertex does not contain the energy
of the $\Delta$ or nucleon, the contribution
from the constant vector mean-fields is eliminated in the calculation 
of the polarization by a simple
change of integration variable.
As the $\Delta$ is unstable, we include a width $\Gamma_\Delta$
by using a complex mass $M^c_\Delta$
\begin{equation}
M_\Delta \rightarrow M^c_\Delta \equiv
M_\Delta-i\,\Gamma_\Delta /2\ ,
\end{equation}
in the denominator of the delta propagator~\cite{tjon}.  This 
prescription
yields similar results as the usual Breit-Wigner folding.  Note, we
have not included pion distortions in Eq.~(\ref{double}).  These may
somewhat reduce the cross section.  The effect
of distortions and a possible $\Delta$-h spreading potential 
remain to be 
investigated.

\section{Results}
\label{sec:result}

In this section, we present numerical results of our calculation.  
We start with the inclusive cross sections and their role in 
atmospheric
neutrinos followed by  a discussion of the coherent pions.
In the case of the inclusive cross section, we 
discuss the case of muon neutrinos, 
but the general features of the results hold for electron neutrinos as 
well.  In addition to the relativistic Fermi gas description of 
the nucleus, the effects of
mesonic mean fields are also considered in the mean-field 
approximation (MFA).  The 
target nucleus is assumed to be $^{16}$O with a 
Fermi momentum $k_F=225$ MeV.  
In MFA, we use the same scalar and vector couplings for nucleon 
and delta
(``universal coupling'').  For $k_F=225$ MeV, the effective masses 
are
\begin{equation}
M^*_\Delta = 931 {\rm MeV}\ ; \quad M^*_N = 638 {\rm MeV}\ ,
\end{equation}
and the vector self-energy is $V = 239$ MeV.

In atmospheric neutrino experiments such as Kamiokande and
IMB~\cite{imb,kamiokande},  the cross section of interest
is $d\sigma /dE_{\bf k'}$, which is
obtained  from Eq.~(\ref{dcross}) by another numerical integration
over the scattering angle of the outgoing lepton.
Fig.~\ref{dsde} shows the cross section
$d\sigma /d E_{\bf k'}$ for $E_\nu=1$ GeV.
Note that the cross sections from the p-h excitation are peaked at a 
large
lepton energy, which corresponds to a small nuclear excitation energy.  
This is
the region where nucleon correlations are important.  The nucleon 
correlations
are modeled using RPA which includes a ($\rho + \pi +g'$)
residual isovector interaction.  The RPA calculation\footnote{Note 
that 
our RPA calculations do not include
$\Delta$-h or mixtures of p-h and $\Delta$-h.} in the mean-field
ground state is also shown in Fig.~\ref{dsde}.
The cross section for stable deltas
vanishes around $E_{\bf k'}=0.7$ MeV, which gives the minimum energy
transfer for delta production.
The curves including the delta width, however,
spread over to higher lepton energies because of the averaging
process.

The mean fields reduce the p-h and the $\Delta$-h responses by 30\%
at the peaks.  However, RPA effects reduce the peak by 50\% at high 
muon
energies.  The reduction of the $\Delta$-h response by mean fields
is interesting in the
Monte Carlo simulation of atmospheric neutrino experiments.
The strength of the $\Delta$-h cross section provides 
the number of pions produced from the $\Delta$ decays.
These pions may cause an uncertainty in determining the lepton flavor 
in the experiment.
Indeed, the Monte Carlo simulation of the
Kamiokande detector~\cite{monte} takes into account the number
of pion events based on a nonrelativistic model by Fogli and
Nardulli~\cite{fogli}.  Their nonrelativistic results are 
qualitatively
reproduced in our free delta calculations.  Thus
mean-field effects are not included in the existing simulations 
and could
provide significant systematic errors.

To get a rough estimate of the reduction in the number of events, 
we fold
$d\sigma /d E_{\bf k'}$ with a simple model of the atmospheric
neutrino flux.  We approximately fit the atmospheric muon neutrino 
flux
with the formula
~\cite{kamiokande} 
\begin{equation}
\phi_{\nu_\mu}(E_\nu) = 220\ E_\nu^{-2.5}
{1 \over {\rm m}^2\ {\rm sr}\ {\rm GeV}\ {\rm sec}}\ ,
\end{equation}
and assume
\begin{equation}
\phi_{{\bar \nu}_\mu}=\phi_{\nu_\mu}=2 \phi_{\nu_e}=2 
\phi_{{\bar\nu}_e}\ .
\end{equation}
Using this flux, we calculate the yield Y in units of events per 
Kton year of
detector exposure and per 100 MeV energy bin,
\begin{equation}
{\rm Y} =
1.1938 \times 10^{40}
\int \phi_{\nu_\mu}(E_{\nu_\mu})
{d\sigma \over d E_{\bf k'}} dE_{\nu_\mu}\label{events}\ ,
\end{equation}
and plot the result in Fig.~\ref{yield} (a) with respect to the 
muon energy
for the p-h and $\Delta$-h excitations.
Note, the $\Delta$-h calculations include the decay width of the 
delta of 115 MeV. 
However, in the case of the inclusive process 
the quantitative results are insensitive to the decay width.  
At low muon energies,
the mean fields in the impulse approximation reduce the total p-h 
events from
33 to 28 while including RPA yields 22 events (in a 100 MeV bin after
one Kton Year of exposure assuming 100 percent efficiency).  
Thus, nuclear structure effects reduce the quasielastic
charged-current events substantially.  
Since the electron events are also reduced by a similar amount by
RPA correlations, this reduction does not directly affect the 
ratio $r$ of
Eq.~(\ref{rat}).  However, it can change the 
{\it energy dependence} of the cross section.

Also, as one can see from Fig.~\ref{yield}~(b), the $\Delta$-h 
contribution
is about 40\% of the RPA calculation at its minimum and, as the 
muon energy
increases, the $\Delta$-h response becomes more important.  
The $\Delta$-h
response is even larger than the RPA response at high muon energies 
(the
ratio is 1.06 at $E_{\bf k'} = 2$ GeV).  Note that the $\Delta$-h 
response
is reduced by  10\% to 25\% due to mean-fields.  Incoherent pion 
events 
should be decreased correspondingly. 
Therefore, the reduction due to the $\Delta$ mean-fields combined 
with
the event reduction from RPA 
should be properly taken into account in the simulation.

Not all deltas decay to $\pi$N in the nuclear medium.  
In fact, in the medium the $\pi$N decay is partially 
suppressed because of Pauli blocking~\cite{wehr90}. However, the 
$\Delta$ 
in the medium has additional channels of pionless decay.
The coincidence experiments of ($p$, $n$) reactions at
the $\Delta$ resonance~\cite{had} showed that a large portion
of deltas in the medium decay without emitting pions.
Furthermore, in theoretical calculations, a value of about
70 MeV has been used as the 2p-2h partial width~\cite{moniz,jain}.
Thus, experimental and theoretical results indicate a
substantial amount of pionless delta decay in the medium.  
This 2p-2h
decay mode of the delta has not been directly included in present
detector simulations.  At best, its effects have been partially
incorporated using a very crude model of pion absorption without 
any explicit reference to $\Delta$ production.

The pionless decay of the delta and the reduction due to the 
mean-fields, suggest that the simulations may substantially
overcount the total number of pions.  
At this moment, we do not know how these effects change the
result for the atmospheric neutrino anomaly.
But the results indicate  that the Monte Carlo simulations in current
atmospheric neutrino experiments can be improved to treat pion events 
more accurately.

Now we present results for coherent pion production in neutrino
scattering.  Coherent pions may be important in atmospheric 
neutrino experiments 
because they could increase the electron-like events and
therefore reduce the ratio $r$ of Eq.~(\ref{rat}).       
A coherent neutral pion produced in neutral-current scattering 
($\nu$, $\nu'$)
immediately decays into two photons and may be counted as 
an electron-like event.   Likewise, a coherent charged pion   
can also be confused with an electron-like event when the 
two charged particles in the final state,  muon (or
electron) and  charged pion, move with small opening angle and make
an isolated electromagnetic shower.  Indeed small-angle scattering 
dominates the coherent cross section.

The single differential cross section 
$d\sigma/dE_\pi$ is obtained by numerically integrating 
Eq.~(\ref{double})
over the scattering solid angle.  Again, we use the free  
width $\Gamma_\Delta = 115~$MeV in our calculation.
Pauli-blocking and additional decay channels for the $\Delta$ in the 
medium
will change this value (and generate a non-trivial 
momentum-dependence).
However, as theoretical in-medium values are still very uncertain,
we will not explore modifications of the width at this point. 
More work  in this direction has to be done in the future.

In the atmospheric neutrino experiments, both $\mu$- and 
$e$-neutrinos 
participate in producing coherent pions.   Furthermore, in contrast 
to the inclusive 
scattering process where the cross sections of antineutrinos are 
strongly suppressed,  neutrino and antineutrino contributions 
to coherent pion production are of the same size.
Combining the events from $\mu$-type antineutrinos as well as $e$-type 
neutrinos and antineutrinos could enhance coherent pion production 
substantially,
and it is of interest to quantitatively compare the coherent 
pion events 
directly with the inclusive charged-current events through
p-h nuclear excitation for a realistic spectrum of neutrinos.  

Using the $e$-neutrino flux ($\phi_{\nu_e}=\phi_{{\bar\nu}_e}$),
we calculate the coherent pion events and compare them with 
electron production in 
$\nu_e$ scattering.  The neutral pion events per kiloton-year are
\begin{equation}
{\rm Y}(\pi^0) =
6 \times1.194 \times 10^{40} \int \phi (E_{\nu_e})
{d\sigma \over d E_\pi} (E_{\nu_e}) dE_{\nu_e}\label{po-events}\ .
\end{equation}
The factor 6 includes the contribution from antineutrinos 
as well as a factor of 2 from $\mu$-neutrinos.  Note, for the 
purposes of illustration, let us assume all coherent
pion events result in electron-like tracks.  The actual 
identification
of these events depends on the details of a detector.
Fig.~\ref{ncohfig}~(a) shows the coherent $\pi^0$
events with and without mean fields and $\Delta$ decay width.  
The electron events from are obtained from
\begin{equation}
{\rm Y}(e) =
1.194 \times 10^{40} \int \phi (E_{\nu_e})
{d\sigma \over d E_{\bf k'}} (E_{\nu_e}) 
dE_{\nu_e}\label{ve-events}\ .
\end{equation}
The differential cross section 
$d\sigma/d E_{\bf k'}$  is calculated
in relativistic impulse and 
in random phase approximation~\cite{hung}.
Note that the horizontal 
axis in Fig.~\ref{ncohfig}~(a) shows either 
the coherent pion or the electron energy.  

Figure~\ref{ncohfig}~(a) also shows that at roughly $E=300$ MeV the 
coherent pion events, neglecting decay width and  
mean fields, are comparable in size to electron events calculated 
in RPA.  
This is certainly interesting because the electron-like events
can be increased by these pions and therefore might help to 
explain the small ratio $r$ 
of atmospheric neutrinos at $E=300$ MeV.  However,
inclusion of mean fields reduces the pion events by a 
factor of 2.   Furthermore,
the $\Delta$ decay width exacerbates the situation: 
the number of events including mean fields and 
$\Delta$ decay width (thin dashed line) are only 10\% of 
the electron events calculated in RPA.

Next, we consider the coherent charged pion events.  
In this case, the
\v{C}erenkov light detector 
records the total energy of pion and muon (or electron) 
assuming they are moving with small opening angle and make an 
isolated 
single track.
Overall energy conservation implies 
\begin{equation}
E_{\nu_\mu}=E_\pi+E_\mu~~~ {\rm or}~~~~ E_{\nu_e}=E_\pi+E_e 
\end{equation}
depending on the incoming neutrino type.

The coherent charged pion events are calculated from 
\begin{eqnarray}
{\rm Y}(\pi^{\pm}) &=&
2 \times\ 1.194 \times 10^{40} \phi (E_{\nu_e}) 
\int^{E_\nu-m_e}_{m_\pi}
{d\sigma \over d E_\pi} dE_\pi\nonumber\ \\
&+&
4 \times\ 1.194 \times 10^{40} \phi (E_{\nu_e}) 
\int^{E_\nu-m_\mu}_{m_\pi}
{d\sigma \over d E_\pi} dE_\pi\
\label{c-events}\ .
\end{eqnarray}
The first term results from  $e$-neutrinos (the factor of 2
includes contribution from antineutrinos) and the second term 
originates from
muon-type neutrinos (with a total flux factor  of 4).

Figure~\ref{ncohfig}~(b) shows the coherent charged pion 
events and the electron events calculated in RPA.
As in the case of $\pi^0$s neglecting mean field and decay width 
of the
$\Delta$ results in cross sections for the coherent pions comparable
or even larger than the quasielastic ones.
However, again mean fields and the
decay width substantially reduce the pion events:  At $E=1$ GeV, the 
full calculation
amounts to only 23\% of 
the quasielastic electron events, but at $E=2$ GeV, 
it increases to 30\%.  
Since the pion events become important as $E$ increases, our
findings provide an important systematic error to 
the interpretation of the recent Kamiokande experiment 
in the multi-GeV energy range~\cite{mult-gev}.

The effect of  neutral and charged coherent pions can be substantial, 
the actual numbers are very sensitive to the delta decay width, 
though.  
Coherent $\pi^0$ production could be  
large at $E=300$ MeV while coherent charged pions may become
important for high visible energies.    
To solidify our results, however, a more elaborate study of the
$\Delta$  properties in the nucleus has to be performed. 

\section{Summary}                                                
\label{sec:con}                                                             

In this work, we have calculated neutrino-nucleus cross sections 
using
a relativistic formalism to investigate possible improvements for
the Monte Carlo simulation of Kamiokande experiment.   
We have found that RPA corrections reduce the charged-current neutrino
events by up to 37\%  for particle-hole excitations. 
Because of this reduction, the $\Delta$-h response could be
important and is found to give significant corrections to 
quasi-elastic
nucleon knock-out processes.  We have found that the relativistic 
mean-fields reduce the $\Delta$-h responses by  10\% to 24\%, which
correspondingly reduce the pion events in atmospheric neutrino 
experiments. 
Combined with the additional channel of the non-pionic decay of the
delta in the medium,  our findings suggest possible uncertainties 
in the present Monte Carlo simulations of detectors and the 
interpretation of the measured data.

We have also discussed coherent pion production in 
neutrino-nucleus scattering.
We have shown that coherent pions can produce a sizeable background
in the atmospheric neutrino measurements. As neutral pions and
collinearly outgoing charged leptons and pions may mimic
electron-type events in the detector, this effect can 
produce uncertainties in determining the flavor ratio of the 
incoming neutrinos.
Coherent pions may provide important systematic errors for the 
recent Kamiokande experiment in the multi-GeV energy 
region~\cite{mult-gev}.
In order to get a clearer quantitative estimate of the influence
of coherent pions, improved calculations should be done and
these should be included in detector simulations
of atmospheric neutrino experiments

\section*{Acknowledgments}
This research was supported in part by the DOE under Grant
No. DE-FG02-87ER-40365
and the NSF under Grant No. NSF-PHY91-08036.
The work of Hunchong Kim was also
supported in part by the Basic Science Research Institute program
of the Korean Ministry of Education through grant no. BSRI-95-2425.
\eject

\appendix
\section*{}

Here we evaluate the nuclear matter polarization, 
$(\Pi_{\rm NM})_{\mu}$,
for coherent pion production.   
The polarization is defined as
\begin{eqnarray}
(\Pi^j_{\rm NM})_{\mu} &=& - i \int{d^4p\over (2\pi)^4}\, 
{\rm Tr} \biggr[\Gamma^j_{\mu\beta}\ S^{\beta\alpha}(p')\ 
\Gamma^j_\alpha (q') \ G^o(p)\biggl]
+ (q_\alpha \rightarrow -q_\alpha)\nonumber \\
&\equiv& \Pi^j_{\mu\nu} (q) q'^\nu\;;~~~ j=(\pi^{\pm}, \pi^0) ,
\end{eqnarray}
where $p'=p+q$.  
The superscript $j$ indicates the polarization for the charged
or neutral pion production.  
The axial-vector ($\Gamma^j_{\mu\nu}$) and pion
$(\Gamma^j_\nu)$ vertices can be found from Ref.~\cite{tjon,ann,towner}.  
The noninteracting nucleon propagator in Hartree approximation reads
\begin{equation}
G^o(p)=(\not\!p + M) {i\pi\over
E_{\bf p}}\delta(p_0-E_{\bf p}){\rm \theta}(k_F-|{\bf p}|)\ .
\end{equation} 
The polarization $\Pi^j_{\mu\nu} (q)$ involves only axial-vector 
vertex and
is given by
\begin{eqnarray}
\Pi^j_{\mu\nu} (q)&=&-i C^j(q^2)
\int{d^4p\over (2\pi)^4}{\rm Tr}
\Biggr[\Gamma_{\mu\beta} S^{\beta\alpha}(p')
\Gamma_{\alpha\nu}\ G^o(p)\Biggl] + 
(q_\alpha \rightarrow -q_\alpha)\label{axial}\ ,
\end{eqnarray}
where 
$\Gamma_{\alpha\nu}=g_{\alpha\nu}-\gamma_\alpha \gamma_\nu / 4$. 
The off-shell term ($\gamma_\alpha \gamma_\nu / 4$) insures 
that the axial
vector vertex satisfies the relation $\gamma^\mu \Gamma_{\mu\nu}=0$.  
The coefficient functions $C^j(q^2)$ are given as
\begin{equation}
\cases{{4  f_{\pi N \Delta} \over 3 \sqrt{2} m_\pi}\ 
{\rm cos} \theta_c\ 
G_A(q^2)\ r_{N\Delta}~~~~{\rm for}~~~~j=\pi^{\pm};  \cr
\cr
{2 f_{\pi N \Delta} \over 3 m_\pi} 
G_A(q^2)\ r_{N\Delta}~~~~~~~~~~~~~{\rm for}~~~~j=\pi^0\ . \cr}
\end{equation}
Here ${\rm \theta}_c$ is the Cabbibo angle and $G_A (q^2)$ is the 
axial-vector form factor for the p-h excitation.  The value
of the axial vector $N\Delta$ transition strength $r_{N\Delta}$ 
is somewhat
model dependent and we choose an intermediate value 2 
in our calculation.

After some algebra, Eq.~(\ref{axial}) becomes
\begin{equation}
\Pi^j_{\mu\nu}(q) =- C^j (q^2) \int_M^{E_F} 
dE_{\bf p}   
\int_{-1}^{1} d \chi \,{ |{\bf p}| \over 8 \pi^2} {T_{\mu\nu}          
\over (p+q)^2 - M_\Delta^2 + i\epsilon } + 
(q_\alpha \rightarrow -q_\alpha)\ .  
\label{co_pol}
\end{equation}
Here $T_{\mu\nu}$ is the result of the Dirac trace,
\begin{eqnarray}
T_{\mu\nu}&=&{1 \over 6 M_\Delta^2}\biggr[3p_\mu p'_\nu p'^2  
+ 3 p_\nu p'_\mu p'^2 + p \cdot p'\ g_{\mu\nu} p'^2 
- 16 p \cdot p' p'_\mu p'_\nu\nonumber \\ 
&&-3 M_\Delta^2 (p_\mu p'_\nu + p_\nu p'_\mu - 5 p\cdot p'g_{\mu\nu}
)\nonumber\\   
&&+ 2 M M_\Delta(9 g_{\mu\nu}\,M^2_\Delta - g_{\mu\nu}p'^2 
 - 8 p'_\mu p'_\nu) \biggl]\ ,
\end{eqnarray}
and $E_F$ is the Fermi energy 
$E_F=\sqrt{k_F^2+M^2}$,  
and the $\chi={\rm cos}\theta$ is the angle between 
{\bf q} and {\bf p}.  
As long as $\Delta$ is stable, the angular
integration can be done analytically while the 
remaining integration over 
$E_{\bf p}$ is performed numerically.  A similar calculation 
in a mean 
field approximation can be done by replacing $M$ and $M_\Delta$ 
by the effective masses. The contribution from vector mean field 
is eliminated
by a change of variable in the energy integration. 
We include the $\Delta$ decay width by replacing the $\Delta$
mass in the denominator of Eq.~(\ref{co_pol}) as
\begin{eqnarray}
M_\Delta \rightarrow M^c_\Delta \equiv 
M_\Delta-i\Gamma/2\ ,
\end{eqnarray}
and the double integrations are done numerically.

\eject
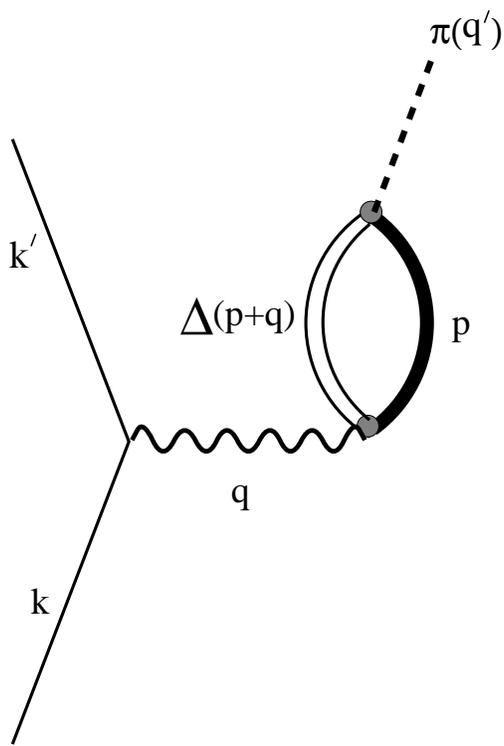
\begin{figure}
\caption {Feynmam diagram for coherent pion production
         through $\Delta$-h excitations.
         In the charged-current reaction, $k'$ is the four momentum
         of the muon or electron, and
         $q$ is the momentum of the $W^{\pm}$ boson.   
         In the neutral-current
         reaction, $k'$ is the momentum of the scattered neutrino 
         and $q$ 
         the momentum of the exchanged $Z^0$.}\label{cohfig}
\end{figure}
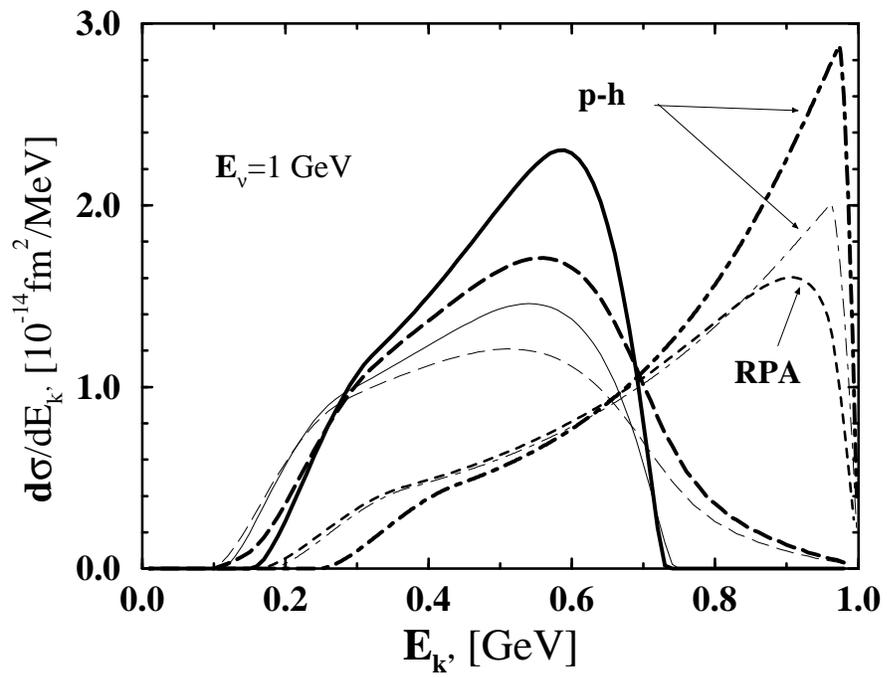
\begin{figure}
\caption{Differential cross section
$d\sigma/dE_{\bf k'}$ as a function of  
muon energy
$E_{\bf k'}$. The thick solid and dashed curves are for the
free $\Delta$-h calculations, without and with decay width,
respectively.  The thick dot-dashed curve is the free p-h calculation.
The corresponding thin curves are for MFA. 
The short dashed curve is the RPA
p-h calculation.}\label{dsde}
\end{figure}
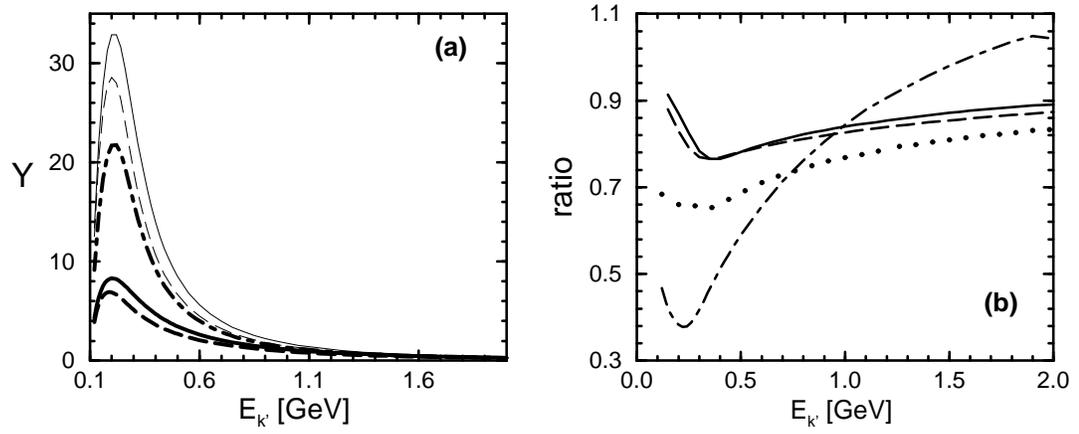
\begin{figure}
\caption{(a) shows events per kiloton-year versus muon energy.  
The bold
solid line represents
the free $\Delta$-h calculation with decay width 115 MeV, and the
bold dashed line is for $\Delta$-h in MFA.  The thin lines
are the corresponding p-h results. Events calculated in RPA are 
given by
the dot-dashed curve.
In figure (b), the solid curve is the ratio of
mean field to free Fermi gas calculations for p-h excitations
and the dashed curve is
the similar ratio of $\Delta$-h events.  The dots gives the ratio of
RPA to free p-h
and the dot-dashed curve is the ratio of free $\Delta$-h over an RPA
calculation of p-h yields. }\label{yield}
\end{figure}
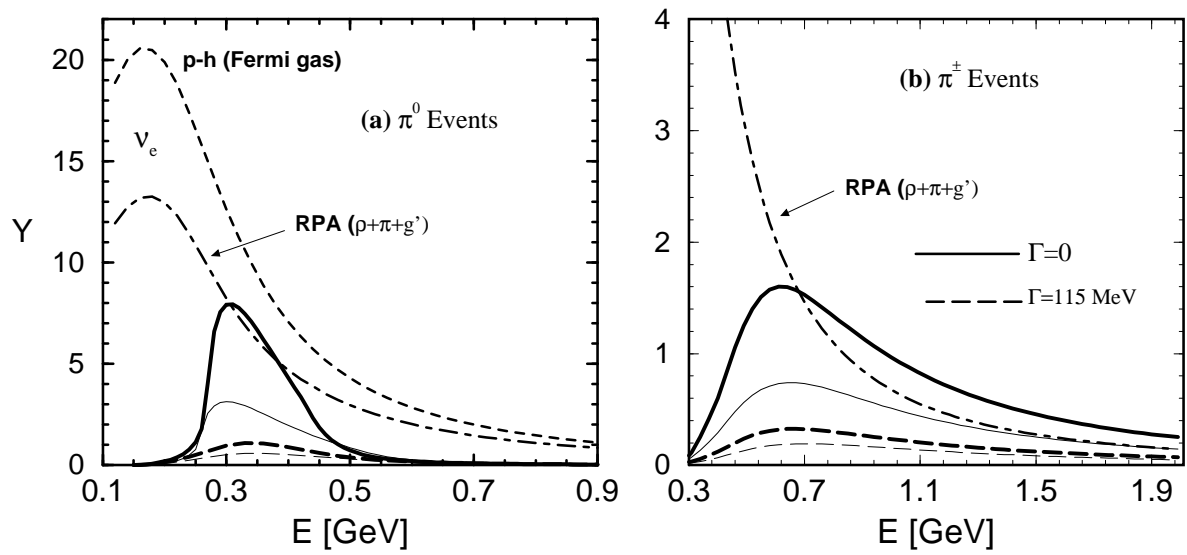
\begin{figure}
\caption{(a) shows electron events calculated in the Fermi gas model 
        (short-dashed curve)
        and RPA (dot-dashed) involving only p-h excitations 
        versus electron energy.  Also shown are the 
        coherent $\pi^0$ events
        calculated for $\Delta$'s with (bold long-dashed) 
        and without decay
        width (bold solid).  The corresponding 
        thin curves are the cases including  mean fields. 
        (b) shows the
similar curves for coherent charged pions.  In this case, 
the events are
plotted versus visible energy as discussed in the text.}
\label{ncohfig}
\end{figure}

\end{document}